\documentclass[adp,fleqn]{w-art}
\usepackage{times}
\usepackage{w-thm}
\usepackage{cite}
\usepackage[]{graphicx}
\begin{document}
%%    The information for the title page will be placed between
%%    \begin{document} and \maketitle. The order of most entries
%%    is determined by the class file and can not be changed by
%%    rearranging them. The maketitle command follows after the
%%    absract.
%%
%%    Most of the following commands will be completed by the publisher.
%%
%%    The copyrightyear is defined in the .clo file as the first argument
%%    of the copyrightinfo command. If the copyrightyear differs from that
%%    value it might be adjusted by the following definition:
%%
%% \renewcommand{\copyrightyear}{2002}% uncomment to change the copyrightyear.
%%
\DOIsuffix{theDOIsuffix}
%%
%% issueinfo for header and copyright line
\Volume{12}
\Issue{1}
\Copyrightissue{01}
\Month{01}
\Year{2003}
%%
%%    First and last pagenumber of the article. If the option
%%    'autolastpage' is set (default) the second argument may be left empty.
\pagespan{3}{}
%%
%%    Dates will be filled in by the publisher. The 'reviseddate' and
%%    'dateposted' (Published online) entry may be left empty.
\Receiveddate{26 July 2007} %\Reviseddate{30 November 2002}
\Accepteddate{31 July 2007 by U. Eckern} \Dateposted{ } %%

%% \pretitle{Editor's Choice}

%% We have a short and a long form for the title. The short form
%% (optional argument) goes into the running head.

\title[Hirsch index for 26 physicists]{A case study of the Hirsch index\\ for 26 non-prominent physicists}

%% Please do not enter footnotes or \inst{}-notes into the optional
%% argument of the author command. The optional argument will go into
%% the header.  If there is only one address the marker \inst{x} may be
%% omitted.

%% Information for the first author.
\author[M. Schreiber]{Michael Schreiber\footnote{E-mail:~\textsf{schreiber@physik.tu-chemnitz.de},
            Phone: +49\,371\,531\,21910,
            Fax: +49\,371\,531\,21919}}
\address {Institut f\"ur Physik, Technische Universit\"at Chemnitz,
09107 Chemnitz, Germany}

\begin{abstract}
The $h$ index was introduced by Hirsch to quantify an individual's
scientific research output. It has been widely used in different
fields to show the relevance of the research work of prominent
scientists. I have worked out 26 practical cases of physicists
which are not so prominent. Therefore this case study should be
more relevant to discuss various features of the Hirsch index
which are interesting or disturbing or both for the more average
situation. In particular, I investigate quantitatively some
pitfalls in the evaluation and the influence of self-citations.
\end{abstract}
%% maketitle must follow the abstract.
\maketitle                   % Produces the title.

%% If there is not enough space inside the running head
%% for all authors including the title you may provide
%% the leftmark in one of the following three forms:

%% \renewcommand{\leftmark}
%% {F. Author: A short title}

%% \renewcommand{\leftmark}
%% {F. Author and S. Author: A short title}

%% \renewcommand{\leftmark}
%% {F. Author et al.: A short title}

%% \tableofcontents  % Produces the table of contents.
\section{Introduction}
Since its introduction by Hirsch \cite{Hirsch} in November 2005
the $h$ index has received a lot of attention, which can be
quantified by the number of already 46 citations of the original
article \cite{Hirsch}, as shown in July 2007 by the Science
Citation Index provided by Thomson ISI in the Web of Science (WoS)
data base. The $h$ index is defined as the highest number of
papers of a scientist that received $h$ or more citations each,
while the other papers have not more than $h$ citations each. It
appears to be an easily computable number which allows one to
estimate the impact of a scientist's cumulative research
contribution because it incorporates both publication quantity and
citation quality. However, as I will discuss below, the
determination of this number from the Science Citation Index is
not as straightforward as it may seem, although the WoS search
allows us to sort papers by number of citations received.\\ The
$h$ index received instantaneous public attention \cite{Bal} and
has been controversially discussed ever since
\cite{Kel,Cro,Raan,Gl2,Bor2}. It has immediately become a subject
of research in informetrics \cite{Bor2,Gl2,Eg2,Eg1,Eg3,Ros}. The
statistical reliability of the $h$ index as compared to other
measures of citation records for discriminating between scientists
has been investigated \cite{Leh}, casting some doubts as to
whether the $h$ index can be used with confidence to distinguish
the scientific achievements in particular, when the number of
publications is relatively small. This point, however, was usually
not a problem in the published evaluations
\cite{Hirsch,Cro,Bat,Gl3} which tried to identify the most highly
cited scientists in various fields, for example, physics and
biological/biomedical sciences \cite{Hirsch}, information science
\cite{Cro,Opp}, physics, chemistry, life sciences and mathematics
\cite{Bat}, ecology and evolution \cite{Kel} and scientometrics
\cite{Gl3}. Very few investigations seem to have been performed
for more average scientists like, for example, the faculty members
of psychological science at Washington University in St. Louis
\cite{Roe}. I have followed the same road and worked out 26
practical cases of not so prominent physicists, thus enabling
observations which should be more common for the data sets of more
average scientists.\\ One shortcoming of the straightforward
evaluation of the citation counts in the WoS data base is the
influence of self-citations which tend to enhance the $h$ index.
Hirsch has argued \cite{Hirsch} that the effect is relatively
small and that only very few, if any, papers with number of
citations just above $h$ would be involved in an appropriate
correction procedure. I have recently shown \cite{MS} that such
self-citations in many cases significantly reduce the $h$ index,
in contrast to Hirsch's expectations, and that the number of
papers for which the self-citation corrections have to be
evaluated is definitely not small, often of the order of 50 \% of
the $h$-defining set of papers and in an extreme case
comprising all highly cited publications of a particular author.\\
Obviously the self-citations do not reflect the impact of the
publication and therefore ideally the self-citations should not be
included in any measure which attempts to estimate the visibility
or impact of a scientist's research.I therefore proposed \cite{MS}
sharpening the $h$ index by excluding the self-citations. For the
26 data sets discussed in the present analysis, I have evaluated
the self-citations not only of the scientist for whom the data set
is analyzed, but also of all co-authors, and I discuss the
respective self-citation corrections for the $h$ index below. On
average, the resulting decrease is significantly larger than
previously reported, for example, for a group of scientists in
ecology and evolution \cite{Kel}, where it was 12.3 \%, or in
information science with 6.6 \%. I attribute this difference to
the assumption that the self-citation corrections are less
significant for highly cited prominent scientists, in agreement
with the observation that Hirsch's $h$ index is reduced by only 10
\%, if self-citation corrections are taken into account \cite{MS}.
This is not so surprising because older studies \cite{Aks,Gl4}
have found that about 25\% of all citations in physics are
self-citations.\\ In the next chapter I will describe the
ascertainment of the data base and its evaluation. First results
of the analysis and the $h$ index are discussed in chapter 3. The
identification of the self-citation corrections is particularized
in chapter 4. The influence of the self-citation corrections and
the sharpened index $h_s$ are analyzed in chapter 5. Finally the
summary and conclusions are given.

\section{Data base}
The subsequent analysis is based on data compiled in January and
February 2007 from the Science Citation Index, which allows
arranging the publication list according to the number of
citations $c(r)$, where $r$ is the rank that the sorting has
attributed to the article. Hirsch's index $h$ is defined by
\begin{equation} \label{equ:H}
h \le c(h) < h+1,\end{equation} or equivalently
\begin{equation} \label{equ:H1}
h = \max_r \big(r \le c(r)\big), \end{equation} which reflects the
verbal definition above, namely that $h$ is given by the highest
number of papers which received $h$ or more citations. In
principle, this value can be directly read off the sorted WoS
list comparing the rank and the number of citations.\\
Out of interest I first checked my own publication list and was
surprised to find an $h$ index of 47. A closer look into the data,
however, showed that the WoS search had found 772 publications out
of which only 270 were my own publications. The reason was
obvious: several other scientists with the same name had
contributed to the list. Such homographs easily distort the
straightforward listing in the WoS data base. Until this analysis
I had not known how common my name is and how many different
subjects appear in the respective WoS search. At first sight, the
subject categories allow for a simple distinction between
homographs. Of course, many of the `wrong' publications in the
list can be excluded easily by using the exclude function in the
WoS thus deselecting, for example, papers in political sciences
and medicine. However, it turned out that there were also a
significant number of publications in physics and physical
chemistry in `my' list that had not been published by myself.
Excluding these by deselecting topics was impossible. Looking at
the titles I could of course identify them, but in many cases it
would have been difficult for another person to decide whether a
paper with such a title could have been written by myself or not.
Most of the WoS data also comprise the authors' affiliations.
Excluding further papers from the list by deselecting certain
institutions can be misleading because affiliations of co-authors
might be taken into account in this way and this could lead to a
wrong exclusion of publications. On the other hand, one can try
positively to select one's own affiliations. I found that in my
own case this procedure does not work well, not only because I
have been working at a number of different institutions during my
career, but also because several of these institutions appear
under a variety of abbreviations in the WoS data. So finally it
appeared to be necessary to compare my own publication list paper
by paper with the ISI list.\\
I then analyzed the publication lists of some colleagues who's
names were not as common as my own so that the investigation was
relatively easy because nearly all papers which were found in the
ISI data base under their names were \underline{really} published
by themselves. Moreover, I had selected colleagues whom I know
rather well so that it was not too difficult to guess whether the
ISI list was more or less correct, even before comparing it paper
by paper with the publication lists. However, the described
selection means a bias in the chosen data sets for my previous
investigation \cite{MS}. Therefore for the present analysis I
decided to extend the number of data sets. It now includes all
scientists who have been working during their research for their
habilitation degree or afterwards as assistants or senior
assistants in my group. Further included are all full and
associate professors from the Institute of Physics at my
university as well as recently retired colleagues. I label the
data sets with superscripts A,B,C,\ldots,Z (this restricted the
number of retired colleagues to four). To include retired
scientists makes sense, because the citation counts and the Hirsch
index can of course increase, even if a person does not publish
anymore. Therefore (also in this aspect) there is scientific life
after retirement.

\begin{vchtable}
\vchcaption{Characteristics for the 26 data sets analyzed in the
present investigation. The data were compiled in January and
February 2007 from the Science Citation Index in the WoS. The
first column labels the data sets, the next column gives a status
of the scientist where 1 indicates an assistant or assistant
professor position, 2 an associate professor position or
equivalent and 3 a full professorship. The following columns show
the Hirsch index $h_{\rm ISI}$ as obtained directly from the ISI
data base without confirming the authorship, the corresponding
total number $n_{\rm ISI}$ of publications in the ISI data base,
the Hirsch index $h$ after substantiating the authorship, and the
number $n$ of publications for which the authorship could be
attributed to the investigated scientist. Finally $n_1$ shows the
number of publications which have been cited at least once, and
$c$(1) the highest citation count for each author.} \label{tab:3}
\begin{tabular}{@{}ccrrrrrr@{}}
\hline
data set & status & $h_{\rm ISI}$ & $n_{\rm ISI}$ & $h$ & $n$ & $n_1$ & $c(1)$ \\
\hline
A & 3 &  39 & 290 & 39 & 290 & 250 & 457\\
B & 3 &  47 & 772 & 27 & 270 & 214 & 182\\
C & 3 &  23 & 126 & 23 & 126 & 103 & 129\\
D & 2 &  20 & 322 & 20 & 322 & 259 &  73\\
E & 3 &  32 & 167 & 19 &  63 &  57 & 279\\
F & 2 &  30 & 450 & 18 & 131 & 107 &  53\\
G & 2 &  17 &  49 & 17 &  49 &  47 &  57\\
H & 3 &  16 &  71 & 16 &  70 &  47 &  70\\
I & 1 &  41 & 544 & 15 &  65 &  53 & 149\\
J & 1 &   3 &  25 & 15 &  51 &  32 & 112\\
K & 2 &  14 &  79 & 14 &  79 &  56 &  55\\
L & 2 &  28 & 309 & 14 &  88 &  67 &  64\\
M & 3 &  15 &  83 & 14 &  70 &  60 & 100\\
N & 2 &  14 &  72 & 14 &  72 &  61 &  55\\
O & 2 &  13 &  76 & 13 &  77 &  66 &  47\\
P & 3 &  13 &  49 & 13 &  47 &  37 & 108\\
Q & 1 &  13 &  91 & 13 &  86 &  59 &  24\\
R & 1 &  14 &  73 & 12 &  46 &  37 &  53\\
S & 2 &  16 & 156 & 12 &  61 &  48 &  40\\
T & 2 &  15 & 134 & 10 &  78 &  56 &  31\\
U & 2 &  10 &  41 & 10 &  44 &  34 &  41\\
V & 3 &  10 &  62 & 10 &  60 &  49 &  79\\
W & 3 &   8 &  49 &  9 &  53 &  37 &  42\\
X & 3 &   8 &  32 &  8 &  35 &  29 & 204\\
Y & 2 &   7 &  28 &  7 &  25 &  19 &  19\\
Z & 2 &   5 &  16 &  5 &  15 &  12 &  25\\
\hline \end{tabular}
\end{vchtable}
\noindent In table 1 I compare the number of papers that are found
in the WoS search in these 26 cases with the number of papers I
could attribute to the authors. In most cases the numbers do not
agree, for nine scientists with a rather common name the WoS
search yields a list which is significantly too large. It is not
surprising that this usually means that the corresponding $h$
index is also unreasonably high. My own case (data set B) is a
good example, because the above mentioned wrong value $h=47$
should really be $h^{\rm B}=27$. Even stronger is the discrepancy
in case I, with $h^{\rm I}=15$ instead of $h=41$. On the other
hand there are also some cases for which the straightforward WoS
search misses a significant number of publications, compare table
1. The reasons were obvious namely the different ways of spelling
names with an umlaut or a suffix such as ``von" in the name. In
one case the original search missed 10 \% of the publications
because the scientist appeared sometimes with one initial and
sometimes with two. In case J not even half the publications were
originally found, because this scientist appears in the WoS with 3
different names due to marriage. Therefore it is very important to
carefully study the data base before evaluation. The problem with
different ways of spelling names is obviously most severe for the
translation or transliteration of names from other alphabets, for
example, for Russian, Chinese and Japanese authors. It is
therefore rather tedious work to establish the data base and it is
usually really necessary to compare with the author's own
publication list in detail. This can also lead to surprising
results. In my present investigation I found two publications in
the WoS in data set M, which were certainly co-authored by the
colleague M, but did not appear in his own publication list. One
of these is even a frequently cited Physical Review Letter, which
enhances the Hirsch index of this colleague. A possible
explanation is that he was the group leader of one of the research
groups involved in the research, but he left the institution
before the publication and has never been made aware of his
honorary co-authorship.\\
The analysis of the various lists confirmed my above-mentioned
reservation concerning the use of the affiliation to attribute
papers to a particular scientist. In the case of my present
university this can be very misleading because of the changes from
Hochschule to Technical University, and University of Technology,
as well as from Karl-Marx-Stadt via Chemnitz-Zwickau to Chemnitz,
and further between faculty, department and institute. In some
cases the WoS data even give only ``inst of phys" as affiliation,
without adding a university
or a city name, and in several cases no affiliation at all.\\
The comparison of the WoS data with the authors' own publication
lists also confirms the often mentioned criticism that the Science
Citation Index comprises only a restricted set of journals.
However, at least in the field of natural sciences, it appears
that all really relevant journals are included and that only some
exotic journals are missing. As publications in these exotic
journals are not likely to be frequently cited there is slight
danger that this restriction influences the evaluation of the $h$
index. Moreover, most citations to papers in such exotic journals
are usually by the authors themselves because only they are aware
of these publications. Therefore after the exclusion of
self-citations the limited WoS coverage of journals should be no
problem at all for the determination of the $h_s$ index.\\ A more
severe restriction is that books and book chapters are not
included in the general search of the WoS data base nor are most
conference proceedings. Thus a significant number of publications
is missing from the subsequent analysis. But this effects all
authors in the same way. Moreover, at least in physics, the
conference proceedings are usually less important publications and
it is therefore not a severe problem when these are not taken into
account. In other fields, with different publication philosophy,
it could be a more severe distortion. For example, an analysis of
the $h$ index and its generalizations applied to computer sciences
showed that in that field conference publications make a
substantial contribution \cite{Sid}.\\ To confirm the expectation
that publications in books are not so relevant for the present
analysis, I have checked for my own case the citations to book
chapters by means of the ``Cited Reference Search" in the WoS and
confirmed that there is no influence on the $h$ index. There are
other data bases which can be evaluated to circumvent the problem
of citations to books, book chapters, and conference proceedings.
For example SCOPUS includes books and conference proceedings, but
suffers from the restriction that cited references are included
only from 1995 onward. Moreover, a strange pattern has been
observed \cite{Jac} as the number of book records in different
years fluctuates strongly. The same odd pattern was found
\cite{Jac} for conference material in SCOPUS. A new and already
popular search engine is Google Scholar, but it yields ridiculous
counts and/or implausible results in simple examples \cite{Jac1}.
The unreasonable number of citations found by Google Scholar can
be traced back to the fact that Google Scholar does not only try
to match papers and citations exactly, but also looks for
approximate matches. In principle this would be a big advantage,
because authors are sometimes rather sloppy with citations
\cite{Jac1}; therefore the WoS search deflates the citation
counts. The number of erroneous citations can be of the order of 5
to 10 \% \cite{Sim} for celebrated papers. But it is more or less
impossible to take such misprints into account, except in very
special cases \cite{Sim}. The approximate matching algorithm by
Google Scholar has been shown \cite{Jac1} to yield nonsense, at
least in some cases.\\ I conclude that in spite of its limited
breadth of covering the Science Citation Index provides the best
data base for the evaluation of the $h$ index. The restrictions
may be more or less unfair for certain scientific fields but
within a field they should have a similar effect, so that
different scientists have the same advantage or suffer from the
same disadvantage. Therefore it is worth noting that for the
present analysis I have selected a rather homogeneous group.

\section{Results of the first analysis: number of papers, citations and the $h$ index}
Comparing the number of papers $n$ of the 26 authors with their
$h$ index in table 1, it is obvious that there is no clear
correlation. The longest publication list belongs to author D,
while scientist E, who is next in the $h$-sorted table 1, would be
shifted to position 15 and G to position 20 regarding the number
of papers. On the other hand, author T would
advance to ninth position, and scientist Q to seventh position.\\
These shifts would be somewhat less drastic, if the numbers $n_1$
of those publications only were taken into account which have been
cited at least once, cp. table1: scientist E would be shifted to
position 11, and G to 17, while author T would advance to 12th and
Q to 10th position. But a clear correlation of $n_1$ with the $h$
index cannot be established, while not surprisingly $n_1$ yields
an arrangement which is similar to that by $n$, with positions
differing by at most four places, on average 1.6 places only.
\\It is interesting to see some associated professors
(D,F,G) high in the $h$-sorted table and also (D,F) regarding the
productivity as measured by the total number of papers. At the
other end of the table the lowest productivity as well as the
lowest $h$ indices are also attributed to associate professors
(Y,Z). It is even more surprising that the next 3 positions from
the bottom of the table are occupied by full professors, with
relatively low numbers of publications as well. On the other hand,
two scientists on the assistant professor level, one of them not
even tenured, appear relatively high in the table at positions I
and J.\\ The maximum number $c$(1) of citations that each of the
26 scientists' publications received, does not yield a clear
correlation with the $h$ index either. As table 1 shows, scientist
X has the third highest citation count and colleague V would also
advance significantly to the ninth position in this respect, thus
overtaking even the above specified most productive author D who
would only appear on the tenth place of the maximum citation-count
list. Likewise, scientists F and G would drop to position 14 and
17, respectively.\\ One might argue, that these observations show
that a single number like the $h$ index or the number of
publications is not sufficient as a measure of the impact of a
particular scientist. Certainly it is better to take into account
more than one such indicator. On the other hand, as already
pointed out by Hirsch \cite{Hirsch} and mentioned in the
introduction, the $h$ index combines both publication quantity and
citation quality. This is confirmed by the above observations
because extreme values of the total number of publications are
smoothed out by the maximum citation counts and vice versa.

\begin{vchfigure}
\includegraphics[width=72mm]{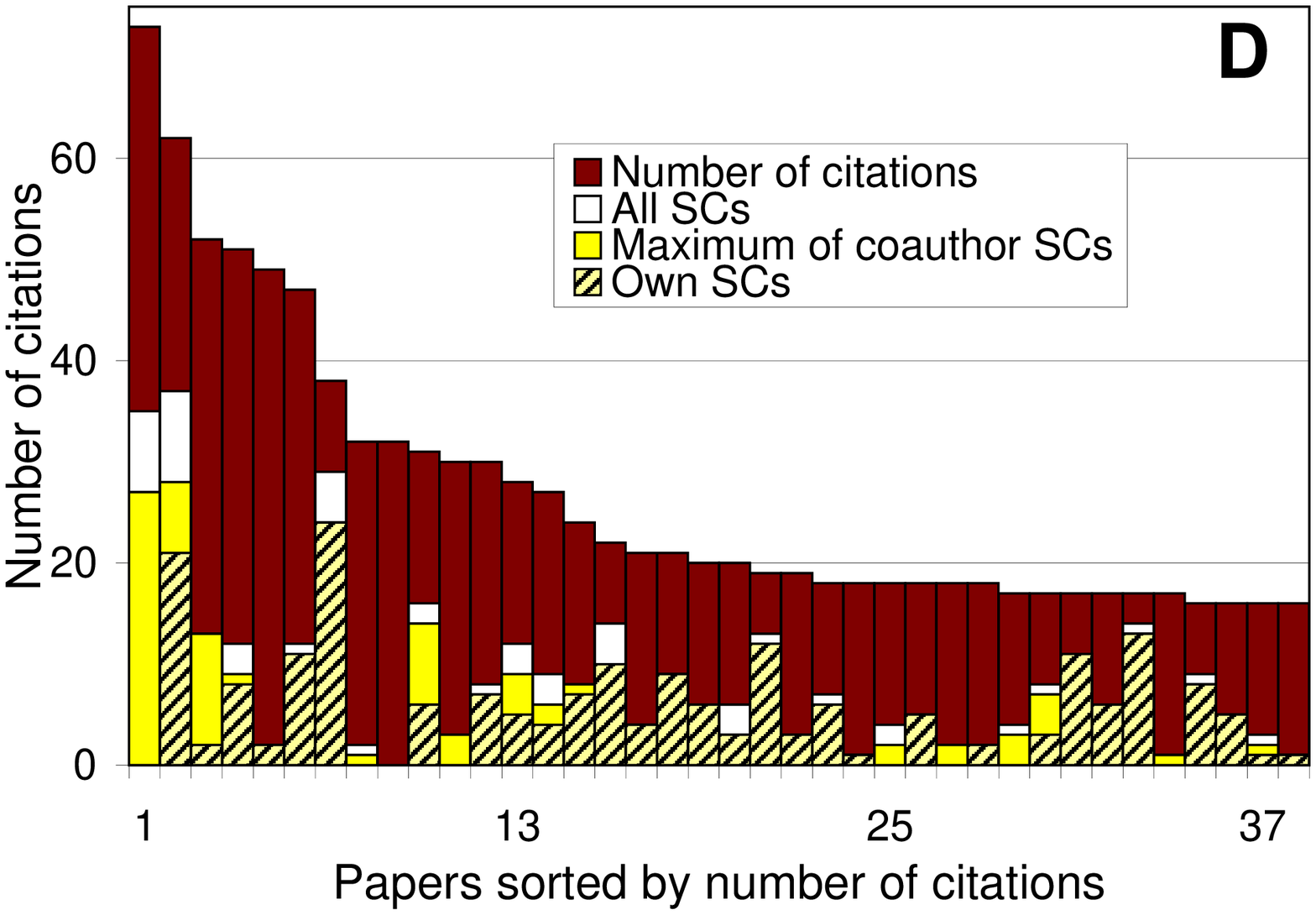}
\hfil
\includegraphics[width=72mm]{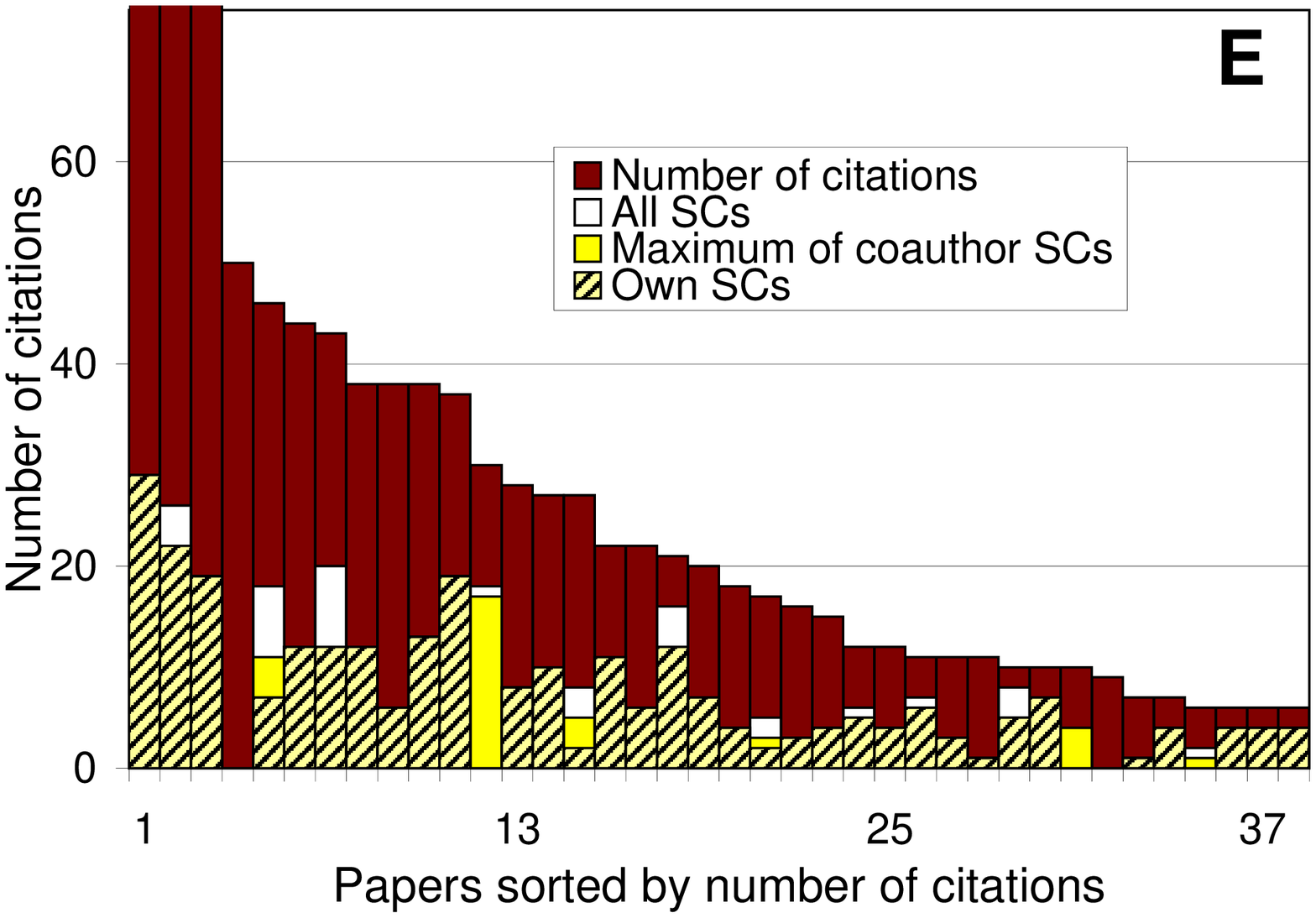}\\
\\
\includegraphics[width=72mm]{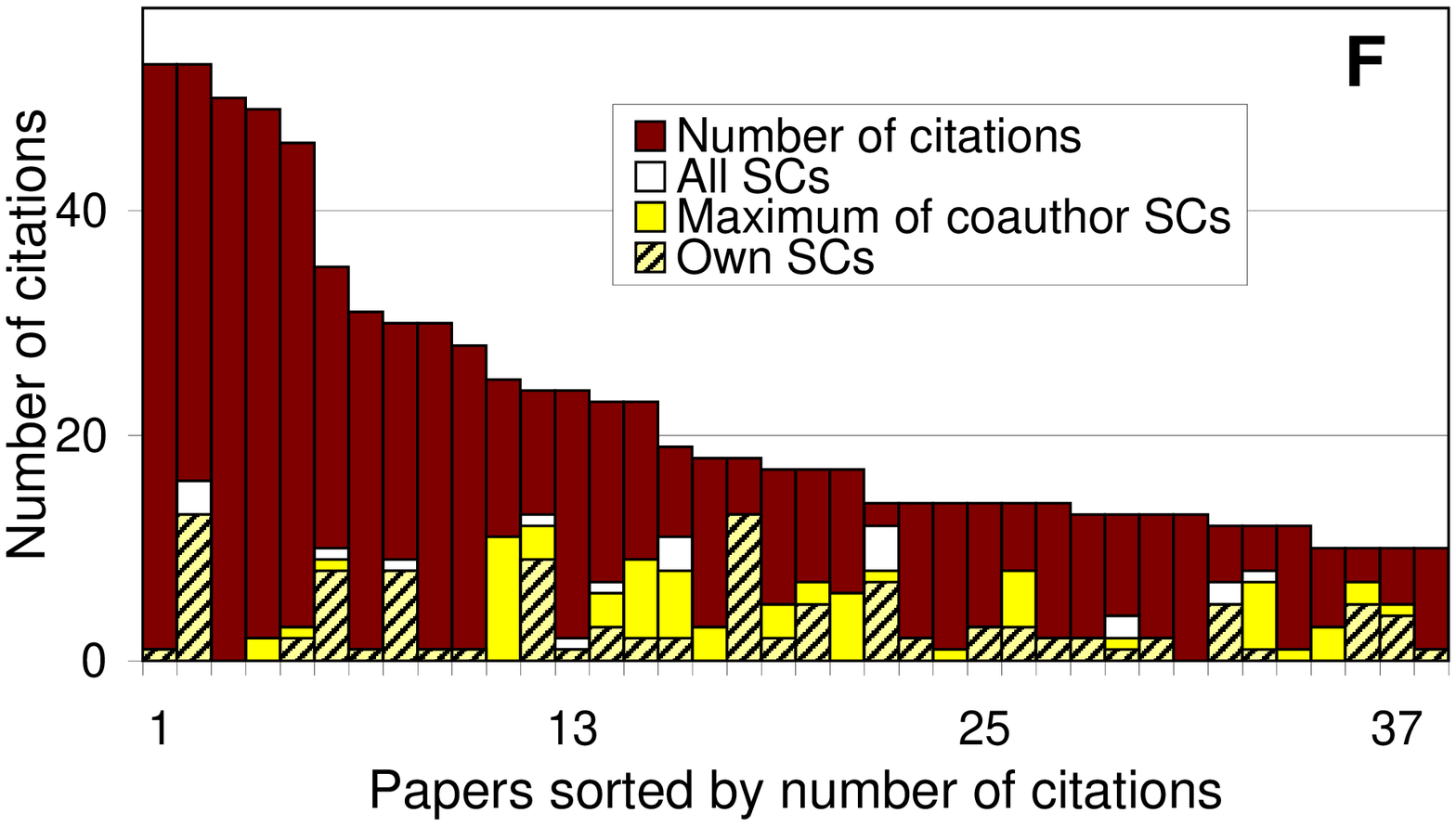}
\hfil
\includegraphics[width=72mm]{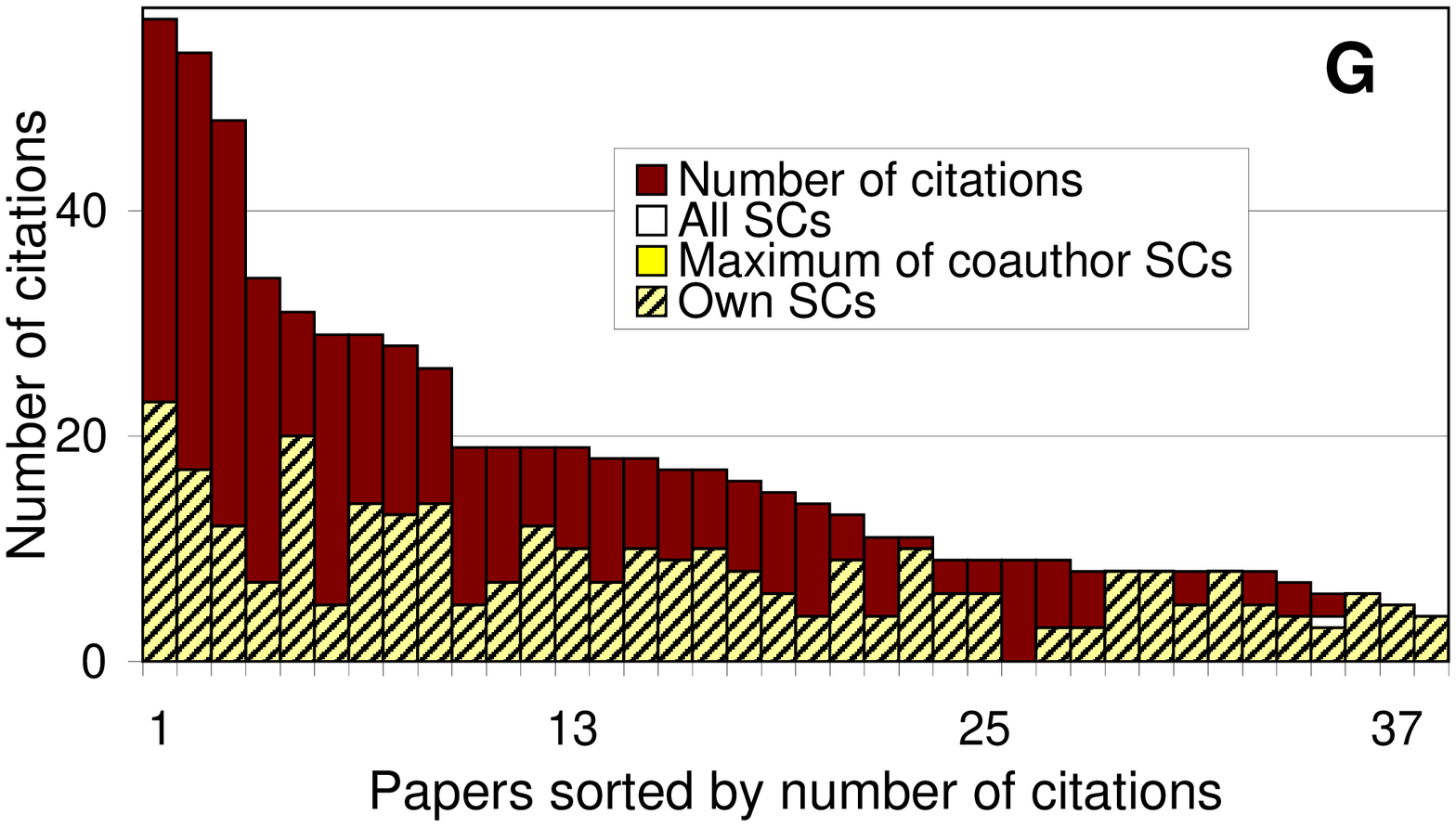}
\caption{Number of citations of the most cited papers in data sets
D,E,F,G, as indicated by the label in the top right corner, (dark
grey/brown), number of the investigated author's own
self-citations (hatched), maximal number of citations by one of
the co-authors including the investigated author (light
grey/yellow), and number of the cumulative self-citations of all
authors (white). As the number of self-citations of the second and
third kind are often equal or not much larger than the
self-citations of the first kind, the respective bars cannot be
distinguished in the plots.} \label{fig:5}
\end{vchfigure}

\section{Identification of the self-citation corrections}
To determine the self-citations one can obtain automatically in
the WoS search the names of up to 100 citing authors for a given
paper and how often these authors cited the respective
publication. In this way one can easily find out how often
somebody has cited his/her own paper. I call these the
self-citation corrections (SCCs) of the first kind and label the
respective quantities with the index $o$ for own SCCs.\\ The
respective data, namely $c(r)-c_o(r)$, are compared with the
citation counts $c(r)$ for data sets D,E,F, and G in figure 1. One
can see that in most cases a significant number of the citations
are self-citations, sometimes more than 50\%. However, for most
papers a substantial number of citations remains when these SCCs
are taken into account. The strongest overall effect can be seen
for data set G, the weakest for data set F. This coincides with
the observation that for these four examples the data set F has
the largest number, namely 9 papers, with zero SCCs (and 8 with
only one SCC and further 8 with only two SCCs) of the first kind
among the 38 papers shown in the plots while there is only one
paper with zero self-citations (and none with one or two
self-citations) among the 38 most cited papers of author G.\\ From
these plots one can expect that the SCCs of the first kind can
only have a small influence on $h^{\rm F}$, but a more important
effect for $h^{\rm G}$, while the changes of $h^{\rm D}$ and
$h^{\rm E}$ should be intermediate because in these cases even
relatively large SCCs of
the first kind still leave a significant number of citations.\\
In any measure of scientific achievement, the self-citations of
the other co-authors should also not be included. Therefore as a
next step I have identified the co-author with the highest number
of citations for a particular publication, again using the ISI
list of citing authors. For long author lists, this procedure
requires a detour via the ``format for print page" which displays
all co-authors, while the WoS summaries show not more than 3
authors. Considering the co-author with the highest citation
count, which might be the author him/herself, I have thus obtained
the respective SCCs which I call the SCCs of the second kind and I
label the respective quantity with the index $c$ for the
co-author SCCs.\\
The respective results, namely $c(r)-c_c(r)$ for the four examples
D,E,F,G are also included in figure 1. One can see that in most
cases the number of self-citations of the second kind is the same
as that of the first kind. For data set G this is always the case,
but this is less surprising when one notes that nearly all of
these papers (41 of 49) are single-author publications.\\ On the
other hand some papers are much more enthusiastically cited by a
co-author than by the here investigated author, most notably paper
1 in data set D, 12 in E, and 11 in F. Nevertheless, the overall
effect is small and thus only little influence on the $h$ index
can be expected.\\ In order to identify the self-citations of all
co-authors, one has to check for all co-author names in the ISI
lists of citing authors. For manuscripts with long author lists
this can be cumbersome especially when some co-authors do not
appear in these lists, which is not so untypical, for example, for
PhD students. As the ISI displays are limited to the set of the
100 most citing authors, in extreme cases a co-author with a
single self-citation might not have been included so that it is
possible, though not likely, that I have committed a slight error
in my analysis in such extreme cases. A simple summation of the
thus obtained self-citation counts of all co-authors of course
overestimates the SCCs, because the self-citations are not simply
additive as two authors of a publication may have written another
article together, citing the first one, which would be counted as
a self-citation for both authors. In order to take these multiple
co-author self-citations properly into account, one has to check
(examine) every citing paper for co-authorship. The WoS search
allows this by checking (ticking off) all co-author names in the
ISI citing author list and then viewing the data, which yields the
number of cumulative self-citations of all co-authors. These I
call the SCCs of the third kind and label the respective
quantities with the index $s$ to denote the sum of all
self-citations. This will yield
the sharpened Hirsch index $h_s$.\\
For the four examples D,E,F,G the obtained results are also
comprised in figure 1, displaying $c(r)-c_s(r)$. The additional
effect in comparison to the previous correction is very small, in
most cases the SCCs of the third kind are not larger than those of
the second kind. There are some exceptions, but overall the
influence on the $h$ index should be small when one compares $h_s$
with $h_c$.\\ Although the observation is trivial, it is important
to note that the ranking of the papers by number of citations is
severely mixed up by the SCCs due to the strongly fluctuating
number of self-citations. To obtain the correct values for the $h$
index with SCCs one has to sort the papers according to the
citation counts after evaluating the SCCs. The resulting Hirsch
indices $h_o$, $h_c$, and $h_s$ are compiled in table 2. In figure
2 the rearranged data are displayed for the 4
examples D,E,F,G.\\
\begin{vchtable} \vchcaption{Influence of
self-citations on the Hirsch index quantified by the indices
$h_o$, $h_c$, and $h_s$ considering the SCCs of the first, second,
and third kind, respectively, as described in the text: $o$
denotes own self-citations, $c$ the maximum of self-citations by
any of the co-authors, and $s$ the sum of all self-citations. Also
given are the values $\hat{r}$, $\underline {r}$, and $\bar{r}$,
i.e. the largest, largest and smallest ranks in the original lists
for which $c(r) > h_s$, $c_s(r)\ge h_s$ and $c_s(r)<h_s$,
respectively, which are utilized in the discussion of the range of
papers for which the SCCs have to be analyzed in order to
determine the sharpened Hirsch index $h_s$. The ratios between the
indices after taking into account the SCCs and the original index
have been calculated from the interpolated values as defined in
Eq.(4). The last column shows the rank which the data set holds
after the list was sorted according to the sharpened index
$\tilde{h}_s$.} \label{tab:3}
\begin{tabular}{@{}crrrrrrrrrrc@{}}
\hline %\vspace {1cm}
\\
data set & $h$ & $h_o$ & $h_c$ & $h_s$ & \hspace {6mm} $\hat{r}$ & $\underline {r}$ & $\bar{r}$ & \hspace {3mm} $\tilde{h}_o$/$\tilde{h}$ & $\tilde{h}_c$/$\tilde{h}$ & $\tilde{h}_s$/$\tilde h$ & order($\tilde{h}_s$)\\
\hline
A & 39 & 36 & 34 & 34 &  46 & 40 & 22 &  0.92 & 0.88 & 0.87 &  1\\
B & 27 & 24 & 23 & 22 &  34 & 28 & 15 &  0.88 & 0.85 & 0.80 &  2\\
C & 23 & 22 & 21 & 20 &  29 & 29 & 13 &  0.93 & 0.91 & 0.87 &  3\\
D & 20 & 17 & 16 & 16 &  34 & 34 &  7 &  0.85 & 0.83 & 0.80 &  4\\
E & 19 & 16 & 15 & 15 &  22 & 17 & 12 &  0.83 & 0.79 & 0.79 &  5\\
F & 18 & 17 & 14 & 14 &  21 & 18 & 11 &  0.94 & 0.78 & 0.78 &  6\\
G & 17 & 11 & 11 & 11 &  21 & 15 & 10 &  0.65 & 0.65 & 0.65 & 13\\
H & 16 & 14 & 13 & 13 &  18 & 15 & 13 &  0.91 & 0.85 & 0.83 &  7\\
I & 15 & 14 & 13 & 12 &  18 & 15 &  9 &  0.91 & 0.87 & 0.82 &  9\\
J & 15 & 14 & 13 & 12 &  15 & 13 & 10 &  0.95 & 0.87 & 0.80 & 10\\
K & 14 & 13 & 13 & 13 &  15 & 15 & 13 &  0.93 & 0.90 & 0.90 &  8\\
L & 14 & 13 & 11 & 10 &  17 & 16 &  6 &  0.90 & 0.76 & 0.69 & 15\\
M & 14 & 11 & 11 & 11 &  17 & 17 & 11 &  0.83 & 0.80 & 0.80 & 12\\
N & 14 & 11 & 11 & 10 &  23 & 15 &  9 &  0.81 & 0.81 & 0.75 & 14\\
O & 13 & 10 & 10 &  9 &  19 & 10 &  7 &  0.79 & 0.75 & 0.71 & 17\\
P & 13 & 13 & 11 & 11 &  13 & 13 &  9 &  1.00 & 0.88 & 0.88 & 11\\
Q & 13 &  7 &  7 &  7 &  23 & 23 &  1 &  0.54 & 0.54 & 0.54 & 22\\
R & 12 & 11 & 10 & 10 &  13 & 11 &  9 &  0.89 & 0.81 & 0.81 & 15\\
S & 12 & 10 &  9 &  9 &  16 & 13 &  8 &  0.83 & 0.79 & 0.79 & 17\\
T & 10 &  9 &  9 &  9 &  18 & 18 &  8 &  0.89 & 0.84 & 0.84 & 19\\
U & 10 &  9 &  8 &  8 &  13 & 13 &  9 &  0.89 & 0.76 & 0.76 & 20\\
V & 10 &  7 &  7 &  7 &  14 & 14 &  7 &  0.76 & 0.68 & 0.68 & 22\\
W &  9 &  8 &  7 &  7 &  13 & 13 &  6 &  0.89 & 0.83 & 0.83 & 21\\
X &  8 &  8 &  7 &  7 &   9 & 10 &  4 &  1.00 & 0.88 & 0.88 & 22\\
Y &  7 &  6 &  6 &  6 &   7 &  7 &  5 &  0.93 & 0.86 & 0.86 & 25\\
Z &  5 &  4 &  4 &  3 &   6 &  3 &  4 &  0.84 & 0.84 & 0.72 & 26\\

\hline \end{tabular}\\

\end{vchtable}
Of course, this reranking should not be restricted to the papers
in the $h$-defining set. Therefore it is necessary to analyze the
citation records somewhat beyond the rank which determines the
Hirsch index. \it {A priori} \rm it is difficult to guess how far
this ``somewhat beyond" leads. To be on the safe side, one should
continue to check all papers as long as $c(r) > h_s$, i.e., the
full citation count is larger than the sharpened index.
The respective rank $\hat{r}$ is included in table 2. \\
Only in retrospect one can ascertain that it is not necessary in
most cases to extend the analysis so far. In table 2, the largest
rank $\underline{r}$ for which $c_s(r)\ge h_s$ is also shown. This
indicates the last paper in the original list, which contributes
to the sharpened index $h_s$. In most cases this rank is larger
than the original $h$ index, which means that this paper did not
belong to the $h$-defining set, but contributes to $h_s$. In 15
out of the 26 cases investigated here it turned out that
$\underline{r}>h$, so one has indeed to extend the analysis
somewhat beyond $h$. Most extreme in this respect are data sets
D,Q, and T, in which cases nearly twice as many papers have to be
analyzed for an accurate determination of $h_s$ as compared to the
calculation of $h$.
\\
Likewise it is not clear \it {a priori} \rm where in the
originally sorted list it is necessary to start the analysis of
the self-citations. One would expect that it is unnecessary to
analyze the papers with very high citation counts because one
expects that for these the exclusion of even a relatively large
number of self-citations is not enough to reduce the remaining
citation counts below the $h$ index. \it{A posteriori} \rm one can
of course determine the rank $\bar{r}$ of the first paper in the
original list, for which $c_s(r)<h_s$, which identifies the first
paper in the original list which drops out of the $h$-defining set
when the SCCs are taken into account. Its rank is also given in
table 2. The values $\bar{r}$ are in most cases (16 of 26) smaller
than two thirds of the $h$ value, in the extreme cases D,L,Q,X
even less or equal to half of the $h$ value. This means that most
of the papers in the $h$-defining set have to be analyzed with
respect to the self-citations. For the most extreme case Q this
includes even the most cited publication.
\\ These observations  are clearly in contradiction to Hirsch's
expectations \cite{Hirsch} that eliminating the self-citations
``would involve only very few if any papers". The analysis also
contradicts Hirsch's statement that ``all self-citations to papers
with less than $h$ citations are irrelevant": When the SCCs were
taken into account, in 15 of 26 cases at least one publication
which did not belong to the $h$-defining set entered the
$h_s$-defining set, because its citation count is in the range
between $h_s$ and $h$, but it has relatively few self-citations. I
again point out that this assessment of how many and which papers
are involved or not involved in the SCCs and thus in sharpening
the $h$ index, has been performed \it{a posteriori} \rm yielding
the range from $\bar{r}$ to $\underline{r}$. The analysis shows
that it is almost impossible to estimate \it {a priori} \rm the
range of citation counts for which the SCCs have to be determined.
I have discussed above the largest rank $\underline{r}$  of a
paper in the original list which entered the $h_s$-defining set.
But in most cases there are further papers in the original list,
for which $c(r)>h_s$, as can be seen in table 2 where often
$\hat{r}>\underline{r}$. These further papers have to be analyzed
with respect to the SCCs because they might enhance $h_s$. The
determination of the SCCs then shows that for these papers
$c_s(r)\le h_s$, so that they do not enhance the $h_s$ value. As a
consequence I cannot give any rule to reduce the effort which is
necessary for the accurate establishment of the SCCs. One just has
to identify the self-citations for all publications until $c(r)
\le h_s$, i.e., up to $\hat{r}$.

\begin{figure}
\includegraphics[width=72mm]{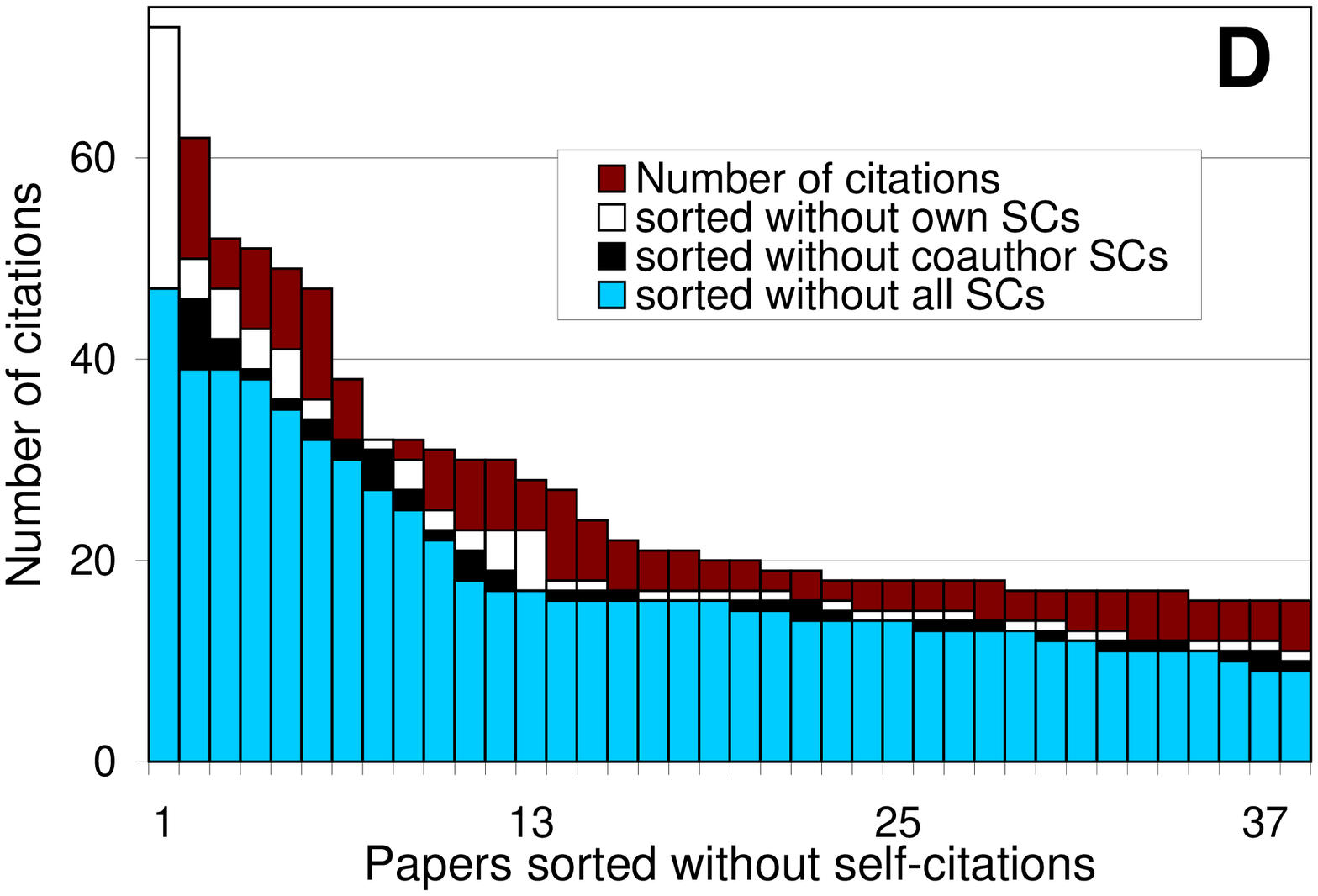}
\hfil
\includegraphics[width=72mm]{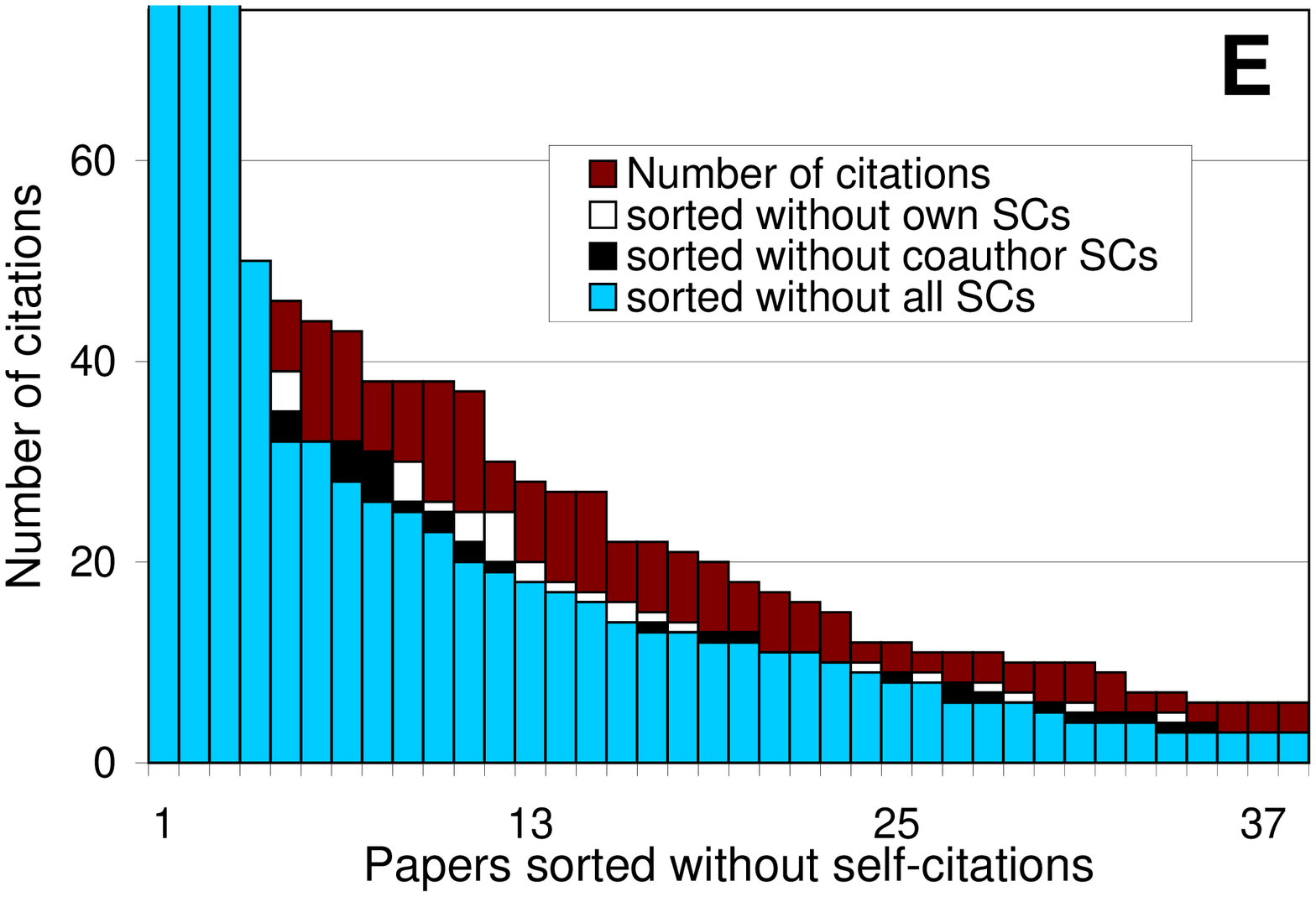}\\
\\
\includegraphics[width=72mm]{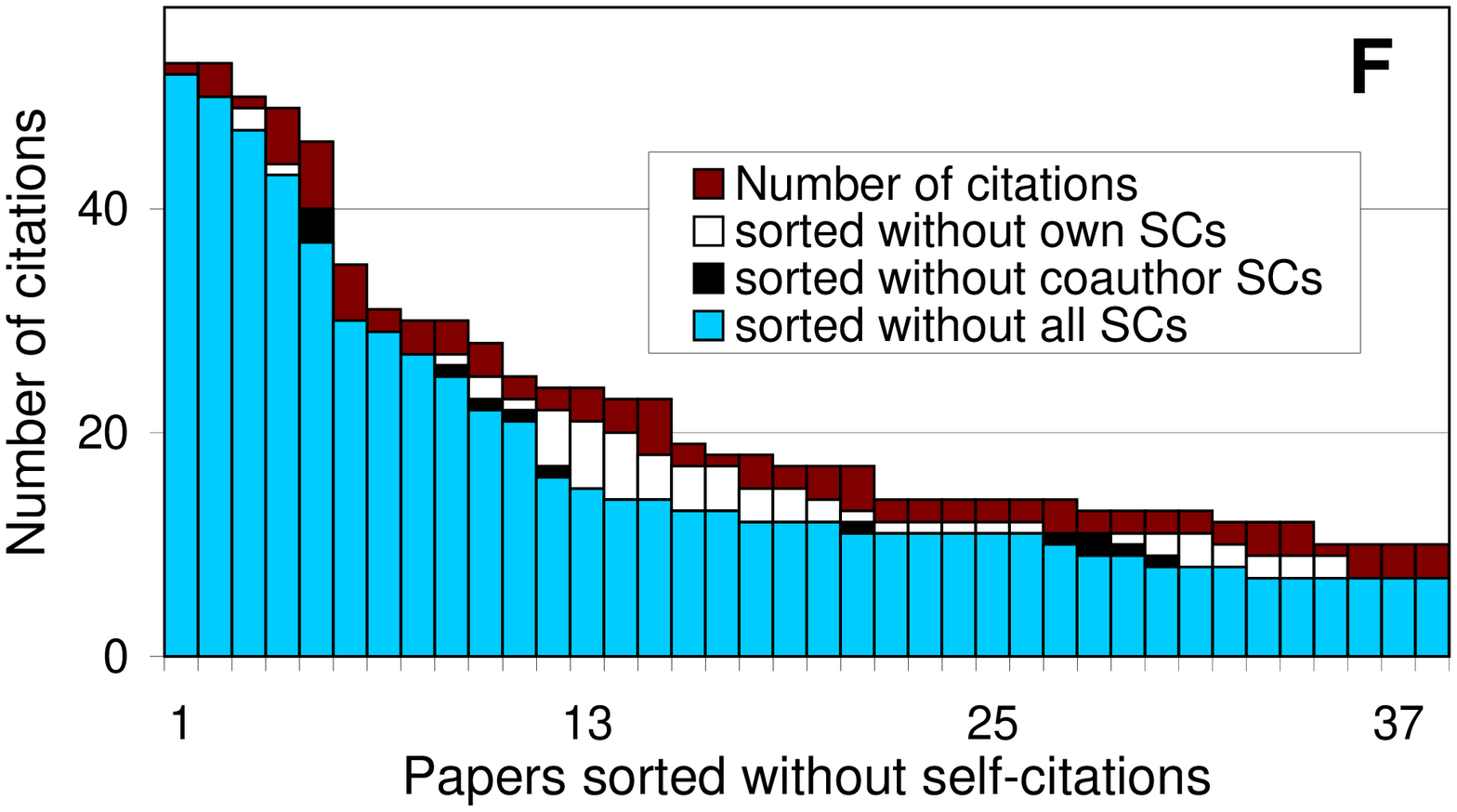}
\hfil
\includegraphics[width=72mm]{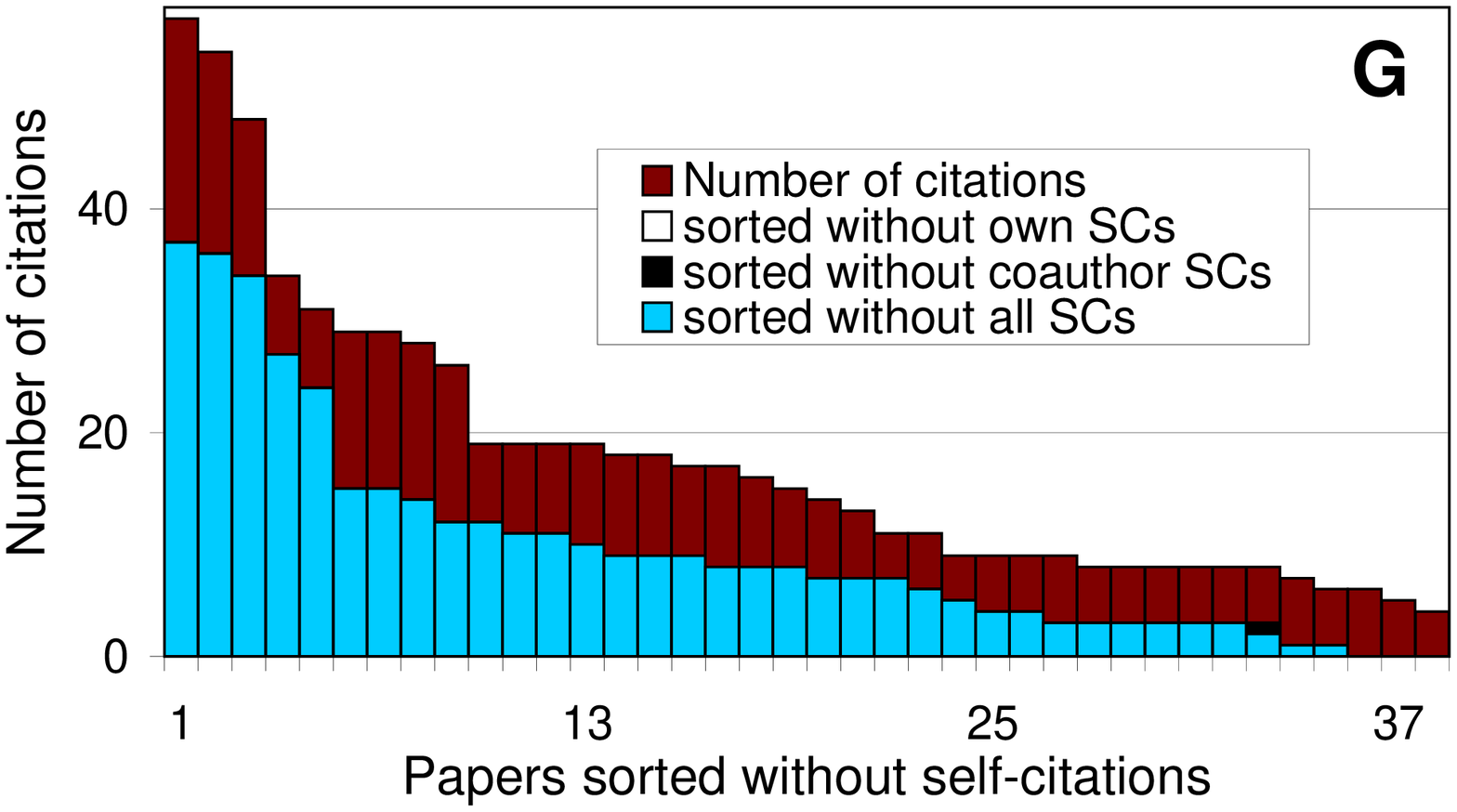}
\caption{Number of citations as in fig.~1, but rearranged after
the exclusion of self-citations. The rearrangement is not
restricted to the papers included in fig.~1 but comprises the full
data set. From top to bottom: total number of citations (dark
grey/brown), without own self-citations, rearranged (white),
without maximal number of any co-author self-citations, rearranged
(black), and without cumulative co-author self-citations,
rearranged (medium grey/blue). As the latter histograms conceal
the previous ones, in particular the columns of 2nd and 3rd kind
often do not show up, because they are not different from the 3rd
and/or 4th kind. }\label{fig:2}
\end{figure}

\section{Results of the second analysis: the sharpened index $h_s$}
The influence of the SCCs on the overall body of citations is
exemplified in figure 2, where the data are put into order by
number of citations after the SCCs of the first kind have been
taken into account, and again resorted after the SCCs of the
second kind were considered, and finally after conclusion of the
SCCs of the third kind. This rearrangement of course usually leads
to the effect that zero SCCs for a particular paper can no longer
be identified (one notable exception can be found for rank 4 in
data set E, because the first 4 papers in this data set need not
be rearranged). The observations made above from the illustrations
of the citation counts of the individual publications in figure 1
can be made more clearly by this way of presentation. In
particular one sees an overall small influence of the SCCs of the
first kind in data set F, but a relatively large reduction by the
SCCs of the second kind especially around the value $h^{\rm
F}=18$. In contrast, for data set G the SCCs of the first kind
dominate, there is no reduction at all by the SCCs of the second
kind and only one small influence at $r_c=33$ by the SCCs of the
third kind.\\ These observations are confirmed by the results
compiled in table 2. A significant reduction of the $h$ index is
found in all cases. But there are interesting differences between
the data sets. As expected from figure 2, the SCCs of the first
kind yield an unusually large reduction from $h^{\rm G}=17$ to
$h_o^{\rm G}=11$. The same absolute reduction occurs for data set
Q, but from a smaller original value $h^{\rm Q}=13$, so that the
relative reduction is even larger. In both cases the SCCs of the
second and the third kind do not further decrease the $h_o$ index.
In my opinion such a strong dominating effect of the own
self-citations is disturbing. There are some other cases
(E,M,N,O,V) in which the effect of the SCCs of the first kind is
also dominating, but less strong, leading to a reduction of
$h-h_o=3$. The same reduction occurs for data sets A,B, and D, but
of course for higher $h$ values so that the relative effect is
smaller. Finally, although for data sets K, T, and Y one finds a
dominating influence, this is counterbalanced by the fact that the
overall reduction is very small, in particular for the data set K
with a higher original $h$ index. It is also worth noting that
there are two data sets (P and X) for which the SCCs of the first
kind yield no reduction at all. This means
that the two colleagues are very unsparing in citing themselves.\\
In most cases the SCCs of the second kind either do not reduce the
$h_o$ values at all (namely in 10 cases) or only by a little (by
$h_c-h_o=1$ in 11 cases) . The strongest absolute effect can be
seen for data set F in agreement with the observations made in the
discussion of figures 1 and 2. A relatively strong effect can also
be found for data sets L and P. In all these cases there are
obviously more enthusiastically self-citing co-authors. \\
Finally, when comparing the sharpened index $h_s$ with the
previous one, $h_c$, one can observe no further reductions in 18
cases, and in the other eight cases a small effect of lowering the
index by $h_s-h_c=1$.
\\
One should, however, be aware of the fact that there exist
somewhat smaller influences of SCCs which are often not reflected
in these values, because Hirsch's definition yields integer
values. It is therefore not surprising, in view of the small
numbers, that quite often there is an agreement between the
derived values. For the analysis of the different kinds of SCCs of
a particular data set, it is therefore helpful to discriminate
somewhat more, which is possible by a piecewise linear
interpolation of the rank-frequency function
\begin{equation}
\label{equ:H} \tilde{c}(x)=c(r)+(x-r) (c(r+1)-c(r))
\end{equation}
between $r$ and
$r+1$, and then calculating the interpolated Hirsch index
$\tilde{h}$ by setting
\begin{equation}
\label{equ:H} \tilde{c}(\tilde{h})=\tilde{h}.
\end{equation}
The respective values are presented in figure 3, which illustrates
the effect of the different SCCs. Especially for this graphical
presentation, I found it better to use the interpolated values
which yield a clearer picture because previously coinciding values
can now be distinguished. One should be aware that the
interpolation will change the index by less than one, if at all.
Therefore the effect is quite small and should not be used to
discriminate between different data sets but only within the same
data set. Nevertheless, I have used the interpolated values
$\tilde{h}$ for the original arrangement of the data sets in
tables 1 and 2, because often the integer values were not
sufficient. For the arrangement in figure 3 I have employed the
interpolated values of the sharpened index $\tilde{h}_s$.\\
\begin{vchfigure}[b]
\sidecaption
%\begin{minipage}{80mm}
  \includegraphics[width=75mm, height=80mm]{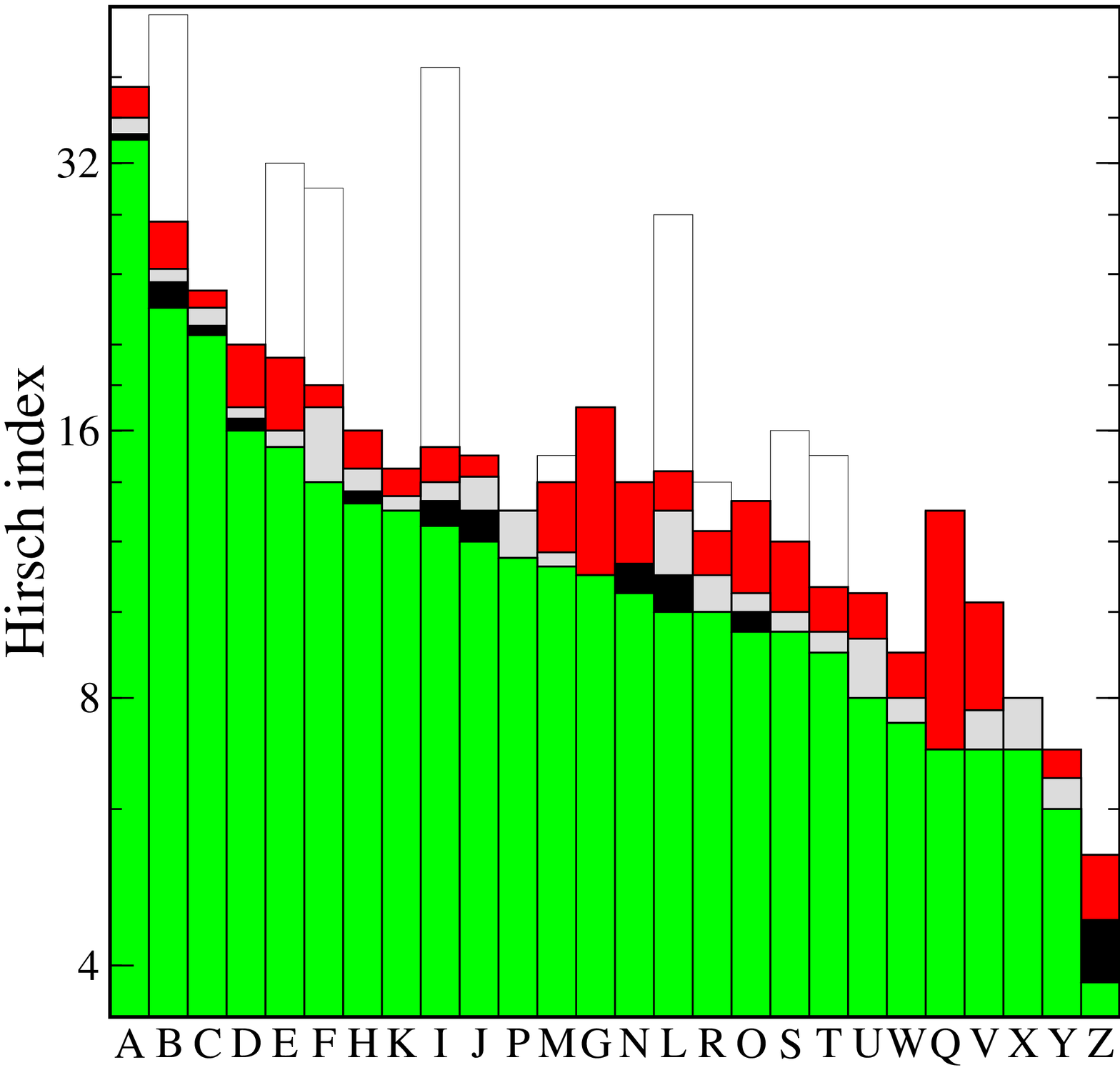}
 % \end{minipage}
\caption{Hirsch indices for the 26 investigated data sets, with
and without SCCs. From top to bottom: brute force index $h_{\rm
ISI}$ read from the ISI data base without confirming the
authorship (white), original Hirsch index $\tilde{h}$ after
substantiating the authorship (dark grey/red), index $\tilde{h}_o$
after exclusion of the author's own self-citations (light grey),
index $\tilde{h}_c$ after exclusion of the maximal number of
self-citations by one of the co-authors including the
investigating author (black), and sharpened index $\tilde{h}_s$
after exclusion of all self-citations (medium grey/green). The
data sets are put into order according to the sharpened index
$\tilde{h}_s$, as indicated at the horizontal axis, where the
letters are not in alphabetical order in contrast to the sequence
in tables 1 and 2 determined by the original index $h$. The latter
histograms conceal the previous ones, so that in particular the
columns of first and fourth kind often do not show up, because
they are not different from the second and/or fifth kind. Note the
logarithmic scale for the $h$ values.}\label{fig:3}
\end{vchfigure}
I have also taken the interpolated values when calculating the
reduction of the indices $h_o$, $h_c$, and $h_s$ in comparison
with the original $h$ index, as shown also in table 2. Using the
interpolated values for the respective ratios demonstrates that in
most cases there is an influence of the SCCs of the second and of
the third kind (now $\tilde{h}_o=\tilde{h}_c$ in 4 cases only, and
$\tilde{h}_c=\tilde{h}_s$ in 15 cases, with
$\tilde{h}_o=\tilde{h}_c=\tilde{h}_s$ in 2 cases, namely G and Q),
although the integer values did not show it. In my opinion, this
interpolation is useful because the integer values in table 2
might otherwise give the impression that, at least in some cases,
it is not worthwhile investigating the SCCs of the second and the
third kind.\\ The observations discussed above for the absolute
values of the indices in table 2 can be made for the relative
values as well. The most extreme cases show up even more clearly
this way: there is no reduction by the SCCs of the first kind for
data sets P and X, while the strongest reduction occurs for data
sets G and Q and this is not reduced any further
by the other SCCs, not even using the interpolated values.\\
Analyzing the (relative) overall reduction from $\tilde{h}$ to
$\tilde{h}_s$ one finds on average a decrease by 21.3\%, which is
significantly larger than the previously reported values. It is
therefore clear that Hirsch's proposition that the effect of
self-citations is small or negligible is not correct. Even if one
restricts the analysis to the SCCs of the first kind, one finds
from the here investigated data sets an average decrease of
13.5\%, and an average decrease of 19.1\%, taking the SCCs of the
second kind into consideration. But more important is the fact
that in several cases the SCCs can help to discriminate between
different data sets, as the relative decrease fluctuates between
46\% and 10\%. The last column in table 2 shows the rank which the
data sets would hold after arranging them according to the
sharpened index $\tilde{h}_s$. Certainly the smaller changes are
not significant but there are also surprisingly large rearranging
effects, which I find quite interesting. In view of the above
discussions, it is less surprising that data sets G and Q fall
back 6 and 5 places, respectively. At the same time the scientist
L is found three positions and O two positions lower on the
rearranged list, whereas several colleagues more forward,
especially colleague P by 5 and K and R by 3 positions. This shows
that the SCCs lead to substantial rearrangement, in contrast to an
investigation of 31 influential information science faculty
members \cite{Cro} for which ``the elimination of self-citations
does not much influence the rank ordering".
\\

\section{Summary and outlook}
This case study of the Hirsch index has been performed for a
relatively homogeneous group of 26 physicists. As the investigated
persons are not so prominent, the results should be more typical
for an average situation than previously reported studies, which
have usually analyzed the publication records of very prominent
persons.\\
It has been demonstrated that it is not straightforward to
determine the data base accurately, a simple WoS search often
leads to a wrong number of publications and consequently usually
to a wrong $h$ index. The main reason is homographs, i.e., authors
who share the same name and initial(s). In figure 3 the bare index
$h_{\rm ISI}$ is also plotted and the exaggeration in comparison
with the Hirsch index $h$ is eye-catching even on the logarithmic
scale. Other difficulties in establishing the data base have also
been discussed and it has become clear that the acquisition effort
in establishing the correct data base can be quite large. It can
therefore be rather misleading and possibly unfair, when strangers
try to determine the Hirsch index by an undiscriminating WoS
search. One way to circumvent this problem would be to ask people
to determine their Hirsch index themselves.\\ The necessary effort
and the possibility of errors are even significantly larger when
one attempts to take the self-citations into account. Again this
is much easier for the person who is under scrutiny.\\ The
significance of these self-citations has been demonstrated in the
present investigation. It was shown that not only the author's own
self-citations have a substantial effect in reducing the Hirsch
index appreciably, but also the self-citations of the co-authors
are usually quite significant and reduce the Hirsch index further.
Unfortunately it turned out that it is not sufficient to analyze
only very few papers with respect to self-citations and that not
even self-citations to papers with less than $h$ citations are
always irrelevant.\\ If the data sets are put into order according
to the sharpened index $h_s$ obtained after all self-citations
have been taken into account, a sometimes drastic shuffling of the
positions has been observed. This corroborates my expectation that
self-citations should be taken into account. It has been argued
\cite{Bal} in favor of the Hirsch index that ``it is hard to
inflate one's own $h$ index, for example by self-citation". The
present investigation shows that this is not so hard, because it
is relatively easy to target one's own publications for which the
citation count is just below the $h$ index and then to cite these
publication on purpose. But when self-citations are excluded it is
much more difficult to inflate the sharpened index $h_s$, although
even this is not impossible,  for example, by cronyism, i.e.,
reciprocal citing. During the preparation of this manuscript I
experienced myself another fascinating way of inflating the
sharpened index, when a referee of one of my manuscripts accepted
in principle the paper for publication but requested the inclusion
of four more references, all of which were (co-)authored by the
same person, even though this scientist had already been cited
five times. It is not unlikely that that (co-)author and the
referee are one and the same person who tried to enhance his/her
$h$ index. I admit that I included the references, as they did
have some connection with my presented research.
\\
The popularity of the Hirsch index is increasing. I believe that
it would be fairer and safer to utilize the sharpened index $h_s$.
Both comprise the information about publication quantity and
citation quality into a single number. The main disadvantages are
the same: firstly the number of co-authors has no influence on the
calculation of the index, and secondly it is not sensitive to one
or several outstandingly high citation counts because, once a
paper has reached the $h$-defining set, it is no longer relevant
whether or not it is further cited. Nevertheless, it is a
reasonable assumption that the Hirsch index will be more
frequently used in the future when assessing the scientific
achievement of scientists for evaluation and promotion purposes.
Let me therefore conclude with a personal note to the esteemed
reader concerning the usage of $h$ ``to reduce a lifetime's work
to a number" \cite{Kel}: I don't like it, you (probably) don't
like it, but let's face it: the $h$ index is here to stay. I
therefore found it worthwhile to point out in the present
presentation the problems one should be aware of when one
determines the $h$ index and/or when one applies it.

%To determine the $g$ index, one has to sum over the number of
%citations up to the rank $r$:
%\begin{equation} \label{equ:H2}s(r) = \sum\limits^r_{r'=1}
%c(r')\end{equation}


\begin{thebibliography}{10}

\bibitem{Hirsch} {J.E.~Hirsch}, {\it An index to quantify an individual's scientific
research output}, {Proc. Natl. Acad. Sci. U.S.A.} {\bf 102}
{(2005)}, {16569 -- 16572}.

\bibitem{Bal} {P.~Ball}, {\it Index aims for fair ranking of scientists}, {Nature} {\bf 436} {(2005)}, {900}.

\bibitem{Kel}
{C.D.~Kelly \and M.D.~Jennions}, {\it The h index and career
assessment by numbers}, {Trends in Ecology and Evolution} {\bf 21}
{(2006)}, {167 -- 170}.

\bibitem{Cro}
{B.~Cronin \and L.~Meho}, {\it Using the h-index to rank
influential information scientists}, {J. Am. Soc. Inf. Sci.
Techn.} {\bf 57} {(2006)}, {1275 -- 1278}.

\bibitem{Raan}
{A.F.J.~van~Raan}, {\it Comparison of the Hirsch-index with
standard bibliometric indicators and with peer judgement for 147
chemistry research groups}, {Scientometrics} {\bf 67} {(2006)},
{491 -- 502}.

\bibitem{Gl2}
{W.~Gl\"anzel}, {\it On the h-index -- A mathematical approach to
a new measure of publication activity and citation impact},
{Scientometrics} {\bf 67} {(2006)}, {315 -- 321}.

\bibitem{Bor2}
{L.~Bornmann \and H.-D. Daniel}, {\it What do we know about the h
index?}, {J. Am. Soc. Inf. Sci. Techn.} {\bf 58} (2007), {1381 --
1385}.

\bibitem{Eg2}
{L.~Egghe}, {\it An improvement of the h-index: the g-index},
{ISSI Newsletter} {\bf 2} {(2006)}, {8 -- 9}.

\bibitem{Eg1}
{L.~Egghe}, {\it Dynamic h-index: the Hirsch index in function of
time}, {J. Am. Soc. Inf. Sci. Techn.} {\bf 58} (2007), {452--454}.

\bibitem{Eg3}
{L.~Egghe \and R.~Rousseau}, {\it An informetric model for the
Hirsch-index}, {Scientometrics} {\bf 69} {(2006)}, {121 -- 129}.

\bibitem{Ros}
{R.~Rousseau}, {\it Simple models and the corresponding h- and
g-index}, {http://eprints.rclis.org/archive}/00006153.

\bibitem{Leh}
{S.~Lehmann, A.D.~Jackson, \and B.~Lautrup}, {\it Measures and
mismeasures of scientific quality}, {arXiv:physics \rm/0512238 and
\it Measures for measures}, Nature {\bf 444} {(2006)}, {1003 --
1004}.

\bibitem{Bat}
{P.D.~Batista, M.G.~Campiteli, O.~Kinouchi, \and A.S.~Martinez},
{\it Is it possible to compare researchers with different
scientific interests?}, {Scientometrics} {\bf 68} {(2006)}, {179
-- 189}.

\bibitem{Gl3}
{W.~Gl\"anzel \and O.~Persson}, {\it H-index for price medalists},
{ISSI Newsletter} {\bf 1} {(2005)}, {15 -- 18}

\bibitem{Opp}
C.~Oppenheim, {\it Using the h-index to rank influential British
researchers in information science and librarianship}, {J. Am.
Soc. Inf. Sci. Techn.} {\bf 58} {(2007)}, {297 -- 301}.

\bibitem{Roe}
{H.L.~Roediger}, {\it The h index in science: A new measure of
scholarly contribution}, {APS Observer} {\bf 19} no. 4 {(2006)}.

\bibitem{MS}
M.~Schreiber, {\it Self-citation corrections for the Hirsch
index}, EPL {\bf 78} (2007), 30002: 1 -- 6.

\bibitem{Aks}
D.W.~Aksnes, {\it A macro-study of self-citation}, Scientometrics
{\bf 56} (2003), 235 -- 246.

\bibitem{Gl4}
W.~Gl\"anzel, B.~Thijs, B.~Schlemmer, {\it A bibliometric approach
to the role of author self-citations in scientific communication},
Scientometrics {\bf 59} (2004), 63 -- 77.

\bibitem{Sid}
A.~Sidiropoulos, D.~Katsaros, Y.~Manolopoulos, {\it Generalized
$h$-index for disclosing latent facts in citation networks},
arXiv:cs.DL/0607066 (2006).

\bibitem{Jac}
P.~Jacso, {\it As we may search - Comparison of major features of
the Web of Science, Scopus, and Google Scholar citation-based and
citation-enhanced databases}, Current Science {\bf 89} (2005),
1537 -- 1547.

\bibitem{Jac1}
P.~Jacso, {\it Deflated, inflated and phantom citation counts},
Online Inform. Rev. {\bf 30} (2006), 297 -- 309.

\bibitem{Sim}
M.V.~Simkin, V.P.~Roychowdhury, {\it Read before you cite!},
Complex Systems {\bf 14} (2003), {269--272}.





\end{thebibliography}
\end{document}